\documentclass[10pt]{article}
\usepackage{epsf} 
\usepackage{amsmath}
\usepackage{amssymb}
\usepackage{epsfig}
\usepackage{latexsym}
\usepackage{amsfonts}
\usepackage{graphicx}%
\usepackage{varioref}
\usepackage{ifthen}
\begin{document}
\parindent 0mm 
\setlength{\parskip}{\baselineskip} 
\thispagestyle{empty}
\pagenumbering{arabic} 
\setcounter{page}{0}
\mbox{ }
\rightline{UCT-TP-275/09}
\newline
\rightline{MZ-TH/09-08}
\newline
\rightline{Revised July 2009}
\newline
\begin{center}
{\large {\bf Confronting QCD with the experimental hadronic spectral functions from tau-decay }}
{\LARGE \footnote{{\LARGE {\footnotesize Supported in part by  NRF (South Africa) and DFG (Germany).}}}}
\end{center}
\vspace{2cm}
\begin{center}
{\bf  C. A. Dominguez} $^{(a)-(b)}$, {\bf N.F. Nasrallah} $^{(c)}$, and {\bf K. Schilcher} $^{(d)}$ \\
\end{center}

\begin{center}
$^{(a)}$Centre for Theoretical Physics \& Astrophysics, University
of Cape Town, Rondebosch 7700, South Africa\\
$^{(b)}$Department of Physics, Stellenbosch University, Stellenbosch 7600, South Africa\\
$^{(c)}$ Faculty of Science, Lebanese University, Tripoli, Lebanon\\
$^{(d)}$Institut f\"{u}r
Physik, Johannes Gutenberg-Universit\"{a}t, Staudingerweg 7, D-55099
Mainz, Germany
\end{center}


\begin{center}
\textbf{Abstract}
\end{center}
\noindent
The (non-strange) vector and  axial-vector spectral functions extracted from  $\tau
$-decay  by the ALEPH collaboration are confronted with QCD in the framework of a Finite Energy  QCD sum rule (FESR) involving a polynomial kernel tuned to suppress the region beyond the kinematical end point where there is no longer data. This effectively allows for a QCD FESR analysis to be performed beyond the  region of the existing data.
Results show excellent agreement between data and perturbative QCD in the remarkably wide energy range $s = 3 - 10 \;\mbox{GeV}^2$, leaving room for a dimension $d$ =4 vacuum condensate consistent with values in the literature. A hypothetical dimension $d$=2 term in the Operator Product Expansion is found to be extremely small, consistent with zero. Fixed Order and Contour Improved perturbation theory are used, with both leading to  similar results within errors. Full consistency is found between vector and axial-vector channel results.
\newpage

\noindent
\section{Introduction}
\noindent

Twenty five years ago \cite{KSMT} it was pointed out that the hadronic decay of the $\tau$-lepton constitutes an ideal laboratory for studying the hadronic weak currents at low and intermediate energies. It was argued that the inclusive character of this decay would make it possible to carry out reliable theoretical calculations in the framework of perturbative QCD (PQCD). Vector and axial-vector final states would be experimentally identified, thus opening new windows for the study of chiral-symmetry. Also, strange and non-strange channels would be separated, allowing for a determination of the strange quark mass. Last, but not not least, the strong coupling
constant would be determined from the total decay rate with impressive
accuracy. The first experimental results from the ARGUS collaboration at DESY \cite{ARGUS} were subsequently used  \cite{DSP}to extract the  values of the QCD vacuum condensates entering the Operator Product Expansion (OPE) and QCD sum rules \cite{QCDSR}. When a high statistics $\tau$-decay experiment   was to become a reality at the ALEPH \cite{ALEPH} and OPAL \cite{OPAL} detectors at LEP, more detailed studies of $\tau$-decay were performed \cite{Braaten}. Following these pioneering analyses, literally hundreds of papers discussing various
theoretical aspects of  $\tau$-decay have been published \cite{taureview}. One of the key issues is that in the limit of vanishing up- and
down-quark masses the  vector and axial-vector spectral functions should become asymptotically equal. This is  not quite yet the case
for the experimental ALEPH spectral functions, even at the highest available
energies. Therefore, PQCD cannot directly be used for
accurate calculations in the vicinity of the positive real s-axis (s being the energy squared). There is, however, sufficient evidence that PQCD and the OPE can be used with impunity for large $|s|$ in the space-like region, as well as in the complex $s$-plane away from the positive real axis. This can be achieved e.g. by using QCD Finite Energy Sum Rules (FESR) with so-called pinched kernels which vanish at the end of the spectrum \cite{MALT}-\cite{DS1}. Some of the
highlights of this procedure are (i) a precise determination of the strong coupling constant \cite{taureview}, \cite{ALPHA1}-\cite{ALPHA2}, (ii) the precocious saturation of the Weinberg sum rules \cite{DS1}, \cite{DS2}, and (iii) the determination of some of the vacuum condensates of the OPE \cite{DS2}-\cite{ASS}. In connection with the latter,
a careful QCD-FESR analysis \cite{DS3} of the  vector and, separately, of the axial-vector (non-strange) channel shows that the values of the gluon condensate $\left\langle
\alpha_{s}G^{2}\right\rangle $ are equal within errors, as expected. This follows, though, only if pinched FESR are employed, as otherwise results from the two channels have opposite signs as observed in \cite{ALEPH}. In addition, as it has been reiterated several times \cite{DS1}, \cite{DS3}, \cite{DS4} the actual values of the condensates are strongly
correlated with the value of $\alpha_{s}(m_{\tau}^{2})$, or alternatively with the value of the QCD scale $\Lambda$. This analysis \cite{DS3} has also found  no clear evidence for a dimension $d=2$ contribution in the OPE, thus confirming results from  earlier analyses  using the ARGUS data \cite{CAD1}, as well as the ALEPH data \cite{DS3}, \cite{DS4}, and leading to values consistent with zero. The conclusions that the gluon condensate from the vector channel is compatible with that from the axial-vector channel, and that a dimension-two term is consistent with zero,  are less apparent from analyses which do not address the convergence problems near the positive real $s$-axis \cite{ALEPH},\cite{NAR}.\\

It is an unfortunate fact that the $\tau$ lepton is not massive enough to allow for a direct confrontation between data and QCD at higher energies. In this paper we attempt to overcome this restriction by introducing
a new FESR involving an integration kernel designed to suppress the hadronic contribution beyond the kinematical end-point of $\tau$-decay, $s_1 = M_\tau^2$, where there is no longer experimental data. We concentrate first on the axial-vector channel, as it involves the well known pion pole. Hence, the weighted integral of the hadronic data up to $s = s_1$, plus that of  PQCD extended beyond this point to $s=s_0 \geq s_1$, can be confronted with the pion decay constant. Results indicate a remarkable and stable agreement between QCD and the ALEPH data over the unusually wide range $s_0 \simeq 3 - 10 \; \mbox{GeV}^2$ , with room for a dimension-four condensate in line with independent determinations. 
A dimension-two term in the OPE, if at all present, turns out to be extremely small. The techniques of Fixed Order PQCD (FOPT) as well as Contour Improved PQCD (CIPT) \cite{CIPT} have been used, with both leading to very similar results within errors. Finally, we study the vector channel along the same lines, except that now there is no scalar meson pole to set a scale. In spite of this, we find perfect agreement with the conclusions drawn from the axial-vector channel. In particular, the value and sign of the gluon condensate from this channel is totally consistent with that from the axial-vector channel.\\
 
It should  be made clear that the purpose of this analysis is definitely not to determine  the pion decay constant nor the vacuum condensates. The former is well known from experiment, and thus supplies a reference scale, and the latter have already been determined with different QCD FESR \cite{DS2}-\cite{ASS}.
Instead, our purpose is to show how to extend a QCD FESR analysis beyond the kinematical region of the existing data. This technique can then be used to confront the data with QCD.
The conclusion from this analysis is that the ALEPH $\tau$-decay data in both the vector and the axial-vector (non-strange) channels is perfectly consistent with QCD.
\noindent
\section{The Finite Energy Sum Rule}
\noindent                                                                           %
We begin by defining the axial-vector current correlator relevant
to $\tau$-decay,
\begin{eqnarray}
\Pi^{\mu\nu}(q^2) &=& i\int d^{4}x \,e^{i q x}\,\left\langle 0|T(A_\mu(x)
A_\nu^{\dagger}(0)|0\right\rangle \nonumber\\ [.2cm]
&=& (-g^{\mu\nu} q^{2} + q^{\mu} q^{\nu}) \Pi_A^{(1)}(q^2) + q^{\mu} q^{\nu}\Pi_A^{(0)} (q^2) \;, 
\end{eqnarray}

where $A^{\mu}(x)= \bar{d}(x) \gamma^{\mu}\gamma_{5}u(x)$. The spectral function used by the ALEPH collaboration, $a_{1,0}(s)$, is related to the imaginary part of the correlator by

\begin{equation}
\operatorname{Im}\Pi_A^{(1,0)}(s)=\frac{1}{2\pi}a_{1,0}(s) \;.
\end{equation}

In this paper we will be concerned first with the axial-vector correlator 
$\Pi_{A}^{(1+0)}(s)$ which we will simply call $\Pi(s)$ for short.  The OPE of $\Pi(s)$ beyond perturbation theory can be written as

\begin{equation}
4\pi^{2} \Pi(Q^{2}) = \sum_{N=0}^{\infty} \frac{1}{Q^{2N}}\;C_{2N}(Q^{2},\mu
^{2})\;\langle \mathcal{O}_{2N}(\mu^{2})\rangle\;,
\end{equation}

where $Q^{2}\equiv-q^{2}$, and $q^{2}$ is large and space-like. The term with $N=0$ above corresponds to PQCD, and the rest of the series parametrize non-perturbative physics. In the standard picture  the term with N=1 is absent from the OPE as there is no gauge invariant operator of dimension $d$ = 2, other than negligible quark mass terms, that could be constructed from the quark and gluon fields. To facilitate
comparison with alternative conventions in the literature,  
absorbing the Wilson coefficients (including radiative corrections) into the
operators in Eq.(3) leads to 
\begin{equation}
\Pi(Q^{2})=\sum_{N=0}^{\infty}\frac{1}{Q^{2N}} \hat{\mathcal{O}}_{2N} \;.
\end{equation}
The spectral function in PQCD to five-loop order, renormalized at a scale $\mu$, is given by

\begin{eqnarray}
4\pi Im \,\Pi(q^{2}) &=& 1 + a_s + a_s^{2}\left(F_3 + \frac{\beta_1}{2} L_\mu \right) 
+  a_s^{3}\left[F_4 + \left(\beta_1 F_3 + \frac{\beta_2}{2}\right) L_\mu + \frac{\beta_1^2}{4}  L_\mu^2  \right] \nonumber \\ [.3cm]
&+& a_s^4 \left[k_3 - \frac{\pi^2}{4}\, 
\beta_1^2 \,F_3 - \frac{5}{24} \pi^2 \beta_1 \beta_2 + \left( \frac{3}{2} \beta_1 F_4 + \beta_2 F_3 + \frac{\beta_3}{2} \right) L_\mu \right.\nonumber \\ [.3cm]
 &+& \left. \frac{\beta_1}{2} \left( \frac{3}{2} \beta_1 F_3 + \frac{5}{4} \beta_2 \right) L_\mu^2 + \frac{\beta_1^3}{8} L_\mu^3 \right] \, ,
\end{eqnarray}
where $L_\mu = \ln (Q^2/\mu^2)$, $a_s \equiv \alpha_s(\mu^2)/\pi$, $\beta_1 = - \frac{1}{2} (11 - \frac{2}{3} \,n_F)$, $\beta_2 = - \frac{1}{8} (102 - \frac{38}{3}\, n_F)$, $\beta_3 = - \frac{1}{32} (\frac{2857}{2} - \frac{5033}{18}\, n_F + \frac{325}{54}\, n_F^2)$, $F_3 = 1.9857 - 0.1153\, n_F$, $F_4 = 18.2427- \frac{\pi^2}{3} (\frac{\beta_1}{2})^2 - 4.2158\, n_F + 0.0862 \,n_F^2$, and the constant $k_3 = 49.076$ has been determined recently \cite{ALPHA2}. The strong coupling constant to five-loop order is \cite{CHET1}

\begin{eqnarray}
\frac{\alpha^{(4)}_{s}(s_{0})}{\pi} &=&
\frac{\alpha^{(1)}_{s}(s_{0})}{\pi}
+ \Biggl (\frac{\alpha^{(1)}_{s}(s_{0})}{\pi}\Biggr )^{2}
\Biggl (\frac{- \beta_{2}}{\beta_{1}} {\rm ln} L \Biggr ) \nonumber \\ [.4cm]
&+& \Biggl (\frac{\alpha^{(1)}_{s}(s_{0})}{\pi}\Biggr )^{3} 
\Biggl (\frac{\beta_{2}^{2}}{\beta_{1}^{2}} ( {\rm ln}^{2} L -
{\rm ln} L -1) + \frac{ \beta_{3}}{\beta_{1}} \Biggr ) \nonumber \\ [.4cm]
&-& \Biggl (\frac{\alpha^{(1)}_{s}(s_{0})}{\pi}\Biggr )^{4}
\Biggl [\frac{\beta_{2}^3}{\beta_{1}^3} ({\rm ln}^3 L
-\frac{5}{2} {\rm ln}^{2} L - 2 {\rm ln} L + \frac{1}{2}) \nonumber \\ [.4cm]
&+& 3 \frac{\beta_2 \beta_3}{\beta_1^2} {\rm ln} L + \frac{b_3}{\beta_1}
\Biggr ] \; , \label{2.9}%
\end{eqnarray}

where
\begin{equation}
\frac{\alpha^{(1)}_{s}(s_{0})}{\pi} \equiv\frac{- 2}{\beta_{1} L}\; ,
\label{2.10}%
\end{equation}

with $L \equiv\mathrm{ln} (s_{0}/\Lambda^{2})$ defines the standard $\overline{MS}$ scale $\Lambda$ to be used here, and

\begin{eqnarray}
b_3&=&\frac{1}{4^4} \Biggl[ \frac{149753}{6} + 3564 \zeta_3
-(\frac{1078361}{162} + \frac{6508}{27} \zeta_3 ) n_F \nonumber \\
&+& (\frac{50065}{162} + \frac{6472}{81} \zeta_3 ) n_F^2 +
 \frac{1093}{729}
n_F^3 \Biggr ] \; , \label{2.11}%
\end{eqnarray}
with $\zeta_{3} = 1.202$. 
Invoking Cauchy's theorem in the complex energy-squared s-plane leads to the standard FESR

\begin{equation}
(-)^N \, C_{2N+2} \, \langle {\cal{O}}_{2N+2} \rangle = 4 \pi^2 \int_0^{s_0} ds\, s^N \, \frac{1}{\pi}\, Im \,\Pi(s) - s_0^{N+1} M_{2N+2}(s_0) \, ,
\end{equation}

where the dimensionless PQCD moments $M_{2N+2}(s_0)$ are given by

\begin{equation}
M_{2N+2}(s_0) = \frac{4\,\pi^2}{s_0^{(N+1)}} \, \int_0^{s_0} ds\,s^N \, \frac{1}{\pi} \, Im \, \Pi(s)|_{PQCD}\;.
\end{equation} 

\begin{figure}[ht]
\begin{center}
\includegraphics[
height=4.0 in,
width=5.0 in
]{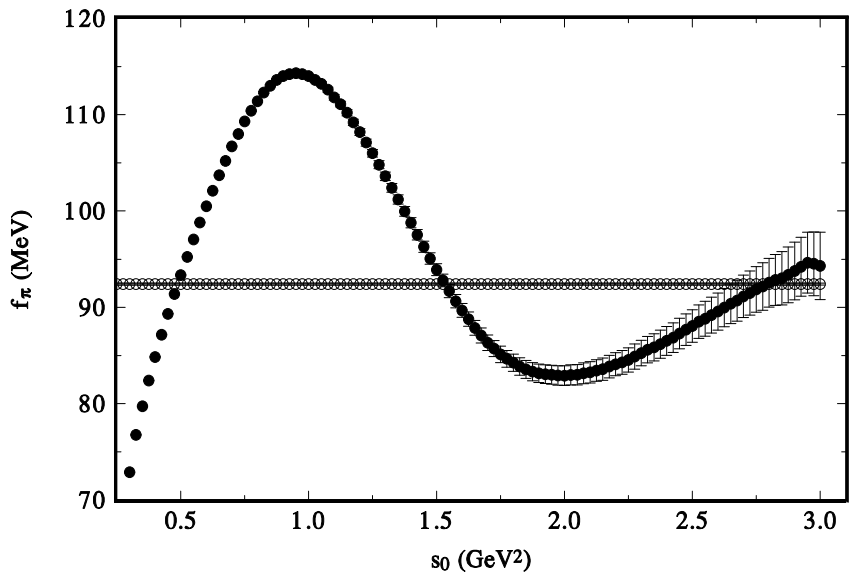}
\end{center}
\caption{Results for $f_\pi$ from the standard  FESR in the axial-vector channel, Eq. (9) with N=0, with no dimension $d=2$ term, and using  CIPT, with $\Lambda = 365 \,\mbox{MeV}$ ($\alpha_s(M_\tau^2)= 0.335$). The straight line  is the experimental value of $f_\pi$.}
\label{FIG1}
\end{figure}

In Fig.1 we show the result for $f_\pi$ from the standard  FESR, Eq.(9) with N=0, using the ALEPH data and CIPT with $\Lambda = 365 \,\mbox{MeV}$ ($\alpha_s(M_\tau^2)= 0.335$), compared with the experimental value $f_\pi = 92.4 \,\pm\, 0.1\, \mbox{MeV}$. Except possibly near the kinematical end-point, the agreement is not particularly encouraging. In fact, it has been known for quite some time that standard FESR in the axial-vector channel do a poor job in reproducing the experimental value of $f_\pi$. Indeed, both the first and the second Weinberg sum rule are not well saturated by the ALEPH data, unless one introduces polynomial integration ({\it pinched}) kernels in the FESR  \cite{MALT}-\cite{DS1}.
This may well be an indication that  quark-hadron duality does not  hold on the real s-axis even at higher energies, except probably very near the kinematical end-point $s_1 = M_\tau^2$. Given the absence of experimental data beyond this point, we propose the introduction in the FESR of an integration kernel, a polynomial $P(s)$,  designed to eliminate the (unknown) hadronic contribution to the integral between $s_1$ and some $s_0 \geq s_1$. The degree of this polynomial $P(s)$, while in principle arbitrary, should not be too high as each additional power of $s$ brings into the FESR an extra higher dimensional condensate with a higher uncertainty. We have found that the optimal degree is in fact the simplest, i.e. a linear function

\begin{equation}
P(s)=1-\frac{2s}{s_{0}+s_{1}} \,,
\end{equation}

so that

\begin{equation}
 \operatorname{constant} \times \int_{s_1}^{s_0} P(s) ds  = 0\,.
\end{equation}

In this case the complete FESR becomes a linear combination of a dimension-two and a dimension-four FESR, which from Eqs.(9) and (11) it is given by

\begin{eqnarray}
2 \, f_\pi^2 &=& - \int_{0}^{s_{1}} ds \, P(s) \, \frac{1}{\pi}\, Im \Pi(s)|_{DATA}
+ \frac{s_0}{4 \pi^2} \left[ M_2(s_0) - \frac{2 s_0}{s_0+s_1} M_4 (s_0) \right] \nonumber \\[.3cm]
&+& \frac{1}{4 \pi^2} \left[ C_2 \langle \mathcal{O}_2 \rangle +\frac{2}{s_0+s_1} C_4 \langle \mathcal{O}_4 \rangle \right] \, + \Delta(s_0)\,,
\end{eqnarray}
where the pion pole has been separated from the data, and the chiral limit is understood. The term $\Delta(s_0)$ is the error being made by assuming that the data is constant in the interval $s_1 - s_0$. While the data is unknown in this region, if one were to assume the onset of PQCD beyond $s = s_1$, then the data would follow the logarithmic fall-off of PQCD. In the next section we will address this issue and quantify this error, which turns out to be very small. Notice that due to the specific form of the integration kernel, $P(s)$, both condensates enter with the same overall sign in the FESR. The cutoff point $s_1$ in Eq.(11) will be chosen as close as possible to the kinematical end-point of the spectrum, in order to include as much data as possible. The sensitivity of the results to this choice will be carefully quantified in the next section.  Another point to be considered is the fact that both $C_2 \langle \mathcal{O}_2\rangle$ and $C_4 \langle \mathcal{O}_4\rangle$ enter in the above FESR. Hence, a study of their correlation will be done in the next section. The FESR in the vector channel follows from Eq.(13) by simply setting its left hand side equal to zero.\\

\noindent
\section{Results}
\noindent
We begin by evaluating the right hand side of the FESR, Eq.(13), in FOPT  and comparing the results with the experimental pion decay constant, $f_\pi = 92.4 \,\pm\, 0.1\, \mbox{MeV}$. We have chosen here, and in the sequel, a cutoff point $s_1 = 2.7125 \, \mbox{GeV}^2$, and defer to the end of this section a discussion about sensitivity to this choice. We recall that in FOPT the strong coupling $\alpha_s(s_0)$ is frozen in Cauchy's contour integral, and the Renormalization Group (RG) is implemented after integration. In CIPT, instead, the coupling is running and the RG is used before integrating. Details of these procedures, as well as explicit formulas for the moments $M_N(s_0)$ may be found e.g. in \cite{DS3}. In order to gauge the impact of the vacuum condensates, we have first set both condensates in Eq. (13) equal to zero and determined $f_\pi$  using the 
value of the strong coupling obtained in FOPT \cite{ALPHA2}, i.e. $\alpha_s(M_\tau^2) = 0.322 \pm 0.004 \pm 0.02$ ($\Lambda = 340 - 347$ MeV). Results are shown in Fig.2. Next, we set $C_2\langle{\cal{O}}_2\rangle$ equal to zero, and  $C_4\langle{\cal{O}}_4\rangle = 0.10 \pm 0.05 \; \mbox{GeV}^4$, according to earlier independent determinations \cite{DS3}, \cite{DS4}-\cite{CAD1}. In this case, the results are shown in Fig.3, which compared with Fig.2 clearly indicates that the contribution of the dimension-four condensate is rather small at these high energies. Even taking into account the rather large error  in $C_4\langle{\cal{O}}_4\rangle$, the agreement between QCD and the data is at an impressive $1 - 2 \%$ level. Turning now to CIPT, and repeating this procedure, but using  $\alpha_s(M_\tau^2) = 0.344 \pm 0.009$ ($\Lambda = 381 \pm 16$ MeV), as determined using CIPT \cite{ALPHA1}, we find the results shown in Fig. 4 (no vacuum condensates), and in Fig. 5 (with condensate values as in FOPT). The experimental value of $f_\pi$ is again reproduced within 1\%, and results from both methods, FOPT and CIPT, are basically identical within errors. Hence, in the sequel we will only show results using CIPT.
 
\begin{figure}[ht]
\begin{center}
\includegraphics[
height=4.0 in,
width=5.0 in
]{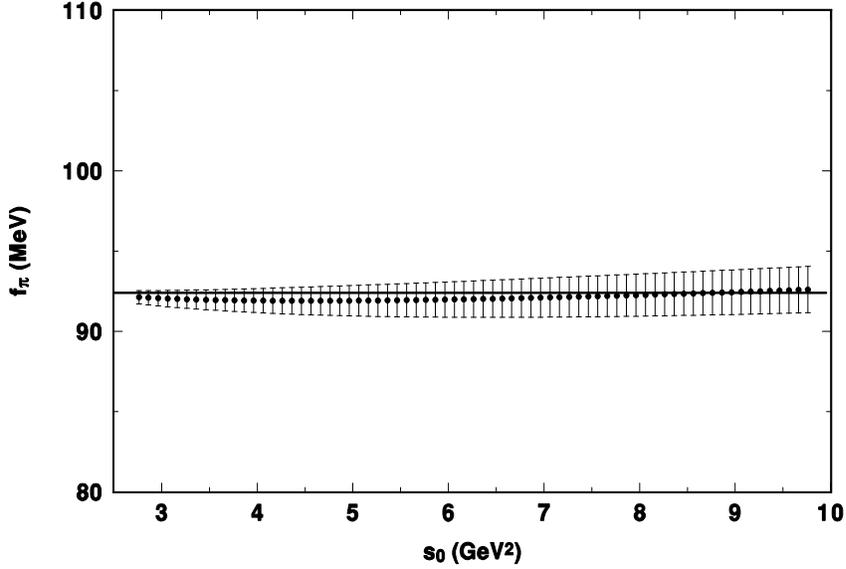}
\end{center}
\caption{Results for $f_\pi$ from the FESR in the axial-vector channel, Eq. (13), with no condensates, 
 and using FOPT with $\Lambda = 343 \,\mbox{MeV}$ ($\alpha_s(M_\tau^2)= 0.322$). The straight line is the experimental value of $f_\pi$.}
\label{FIG2}
\end{figure}

\begin{figure}[ht]
\begin{center}
\includegraphics[
height=4.0 in,
width=5.0 in
]{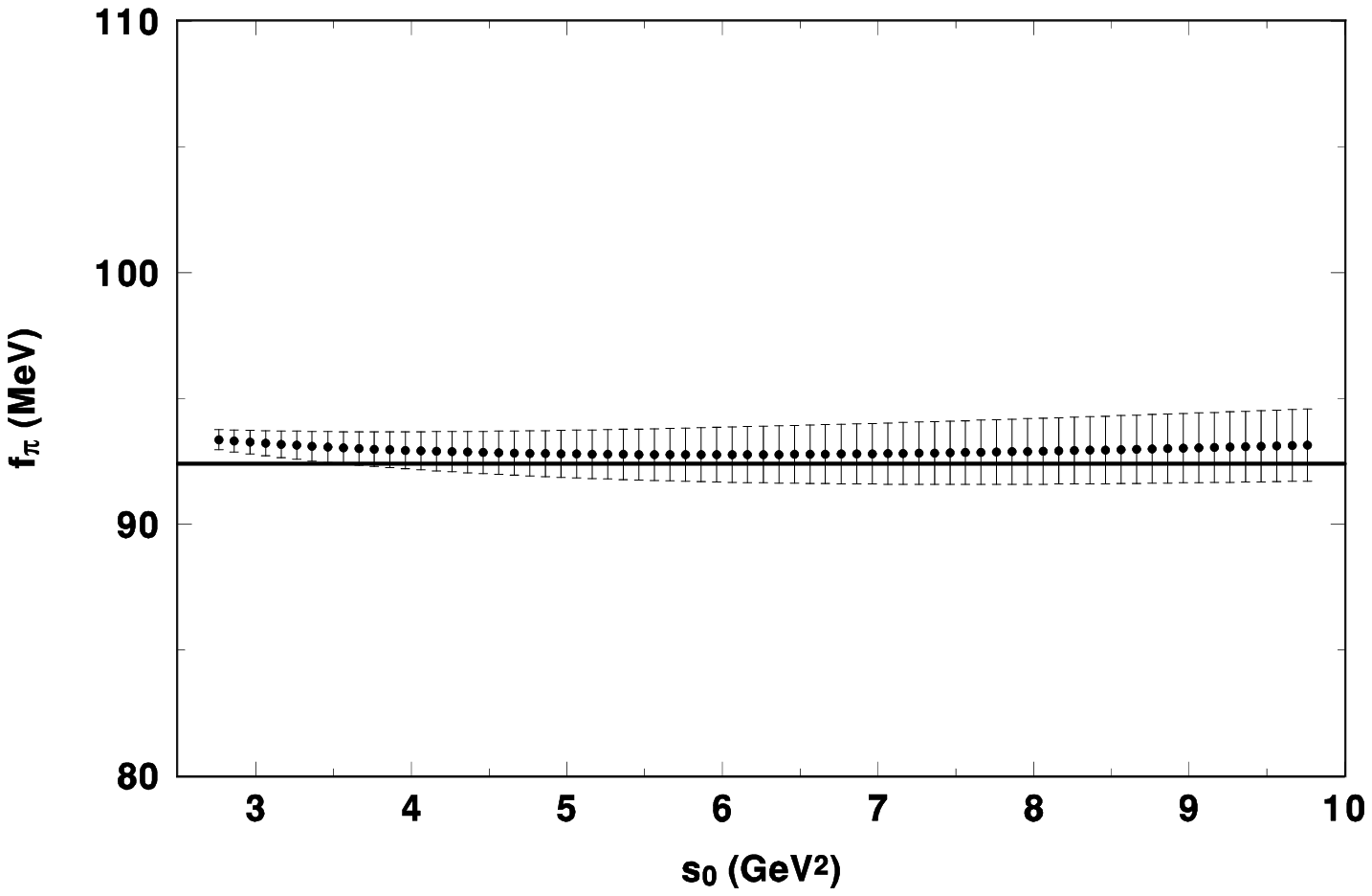}
\end{center}
\caption{Results for $f_\pi$ from the FESR in the axial-vector channel, Eq. (13), with $C_2\langle{\cal{O}}_2\rangle = 0$,
 $C_4\langle{\cal{O}}_4\rangle = 0.05 \; \mbox{GeV}^4$ and using FOPT with $\Lambda = 343 \,\mbox{MeV}$ ($\alpha_s(M_\tau^2)= 0.322$). The straight line is the experimental value of $f_\pi$.}
\label{FIG3}
\end{figure}
 
\begin{figure}[ht]
\begin{center}
\includegraphics[
height=4.0 in,
width=5.0 in
]{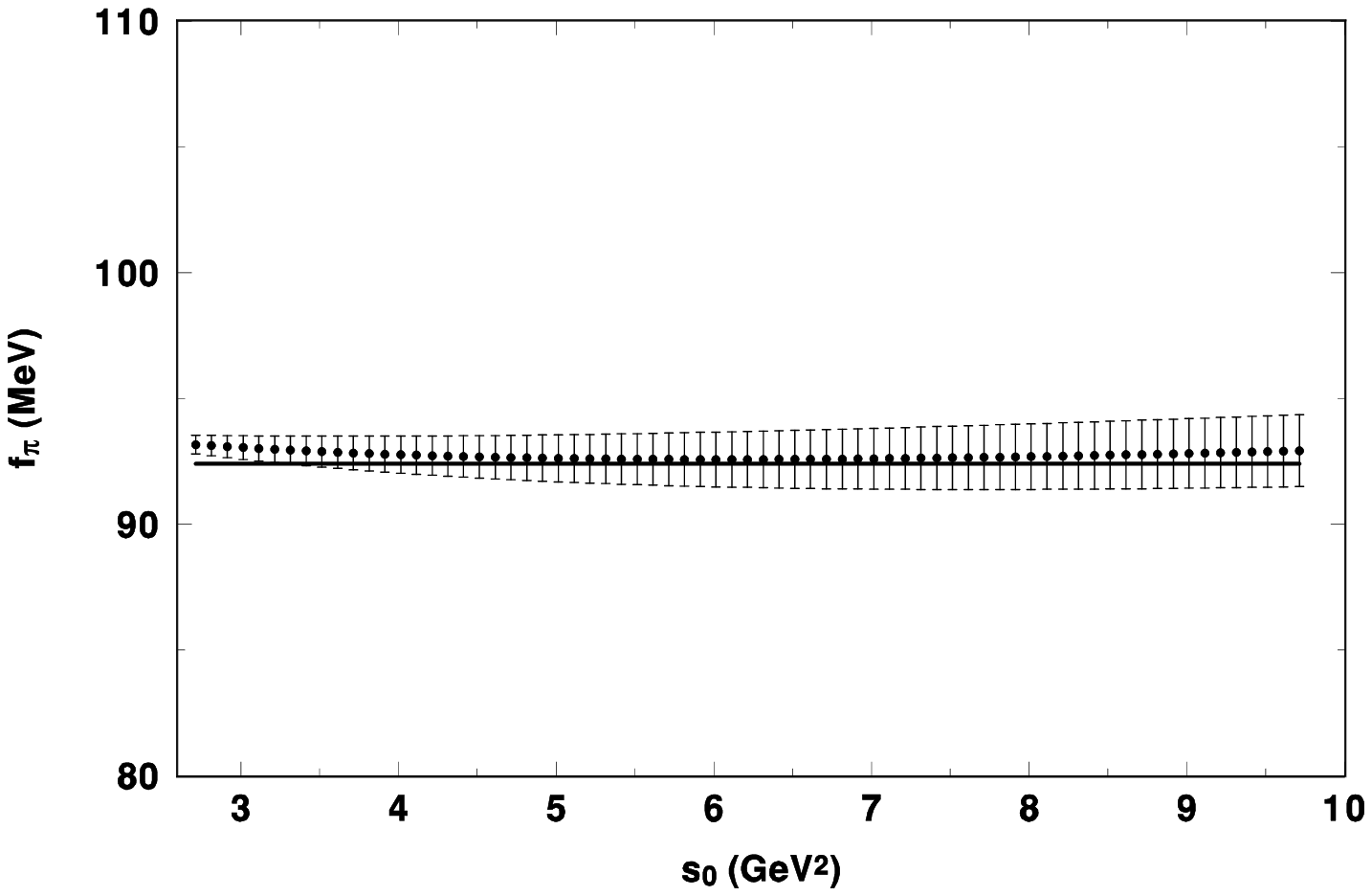}
\end{center}
\caption{Results for $f_\pi$ from the FESR in the axial-vector channel, Eq. (13), with no condensates and using CIPT with $\Lambda = 380 \,\mbox{MeV}$ ($\alpha_s(M_\tau^2)= 0.343$). The straight line is the experimental value of $f_\pi$.}
\label{FIG4}
\end{figure}

\begin{figure}[ht]
\begin{center}
\includegraphics[
height=4.0 in,
width=5.0 in
]{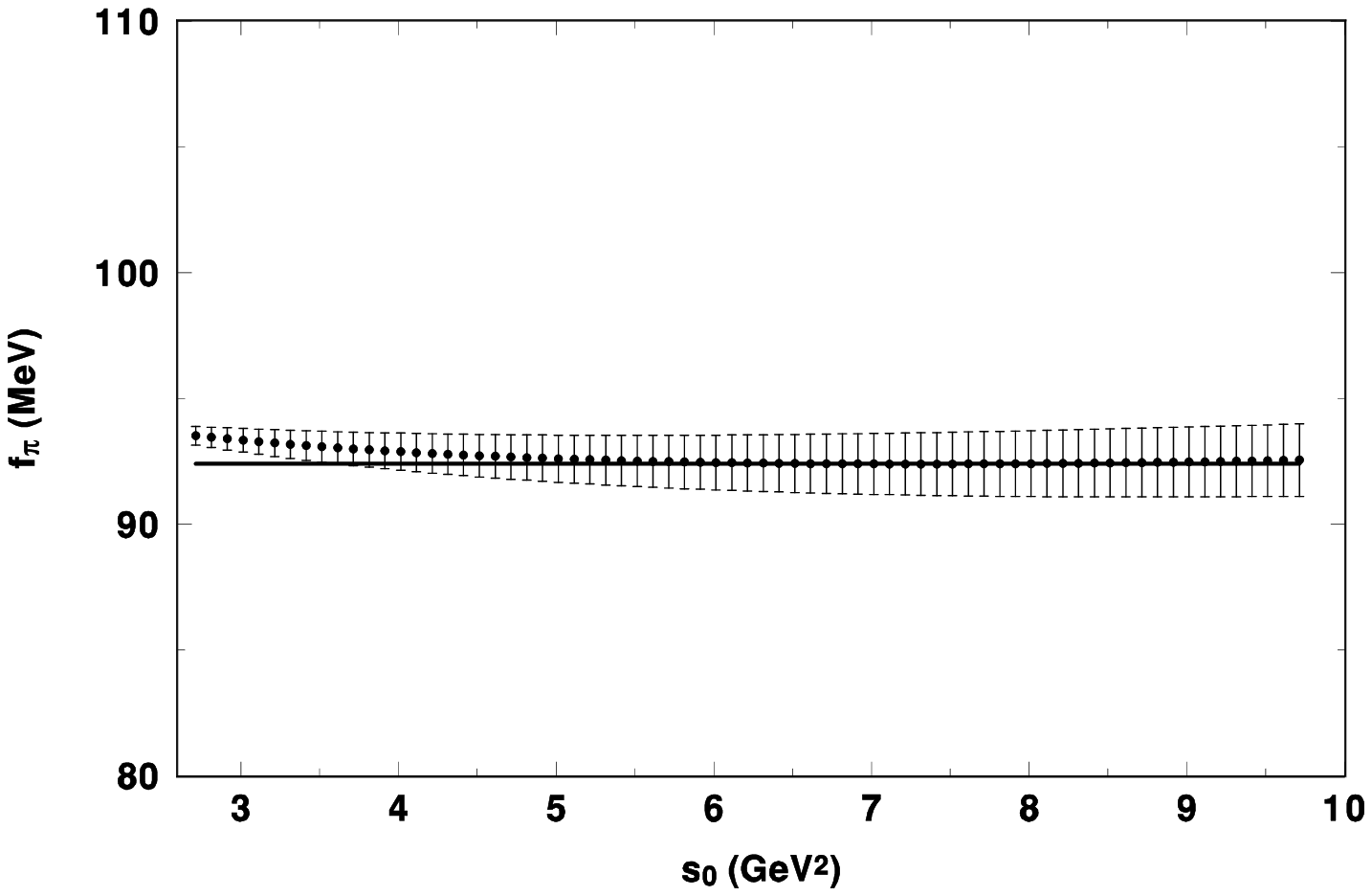}
\end{center}
\caption{Results for $f_\pi$ from the FESR in the axial-vector channel, Eq. (13), with $C_2\langle{\cal{O}}_2\rangle =0$, $C_4\langle{\cal{O}}_4\rangle = 0.05 \,\mbox{GeV}^2$, , and using  CIPT  with $\Lambda = 365 \,\mbox{MeV}$ ($\alpha_s(M_\tau^2)= 0.335$). The straight line  is the experimental value of $f_\pi$.}
\label{FIG5}
\end{figure}

In order to study the presence of a dimension $d=2$ term we shall follow two strategies. First, we allow for changes in the resulting pion decay constant of the order of, say,  $1 \,\mbox{MeV}$, i.e. a 1 \% variation, and introduce a non-zero $C_2\langle{\cal{O}}_2\rangle$ with $\Lambda$ and 
$C_4\langle{\cal{O}}_4\rangle$ fixed. This leads to the maximum value $C_2\langle{\cal{O}}_2\rangle \, \simeq 0.8 \, \times \, 10^{-4} \,\mbox{GeV}^2$, well consistent with zero. Second, we allow $C_4\langle{\cal{O}}_4\rangle$ to change
with $\Lambda$ fixed, and introduce a dimension $d=2$ term in the FESR constraining the result to reproduce the experimental value of $f_\pi$. In this case $C_2\langle{\cal{O}}_2\rangle$ remains definitely  small, e.g. for $C_4\langle{\cal{O}}_4\rangle = 0.1 \, \mbox{GeV}^4$, and $\Lambda = 397 \, \mbox{MeV}$ ($\alpha_s(M_\tau^2)= 0.353$) we find $C_2\langle{\cal{O}}_2\rangle \simeq - \, 0.03 \, \mbox{GeV}^2$. Reducing $\Lambda$ and $C_4\langle{\cal{O}}_4\rangle$ can change the sign of the dimension $d=2$ term, which can reach a maximum value
$C_2\langle{\cal{O}}_2\rangle \simeq 0.005 \, \mbox{GeV}^2$, for 
$\Lambda = 365 \, \mbox{MeV}$ ($\alpha_s(M_\tau^2)= 0.335$), and $C_4\langle{\cal{O}}_4\rangle \simeq 0.05 \, \mbox{GeV}^4$.
The presence of this term in the FESR turns out to be detrimental to the stability region, which tends to become narrower as it approaches the experimental value of $f_\pi$ at much higher energies.  Our result can be compared with a model dependent estimate \cite{NAR}, of substantially different  order of magnitude,
$C_2\langle{\cal{O}}_2\rangle \simeq \, +\,(0.17 - 0.42) \, \mbox{GeV}^2$. 
Clearly, while our analysis does not categorically rule out a dimension $d=2$ term in the OPE, the results are consistent with earlier conclusions that there is no compelling evidence for such a term \cite{DS3}, \cite{DS4}-\cite{CAD1}.\\

\begin{figure}[ht]
\begin{center}
\includegraphics[
height=4.0 in,
width=5.0 in
]{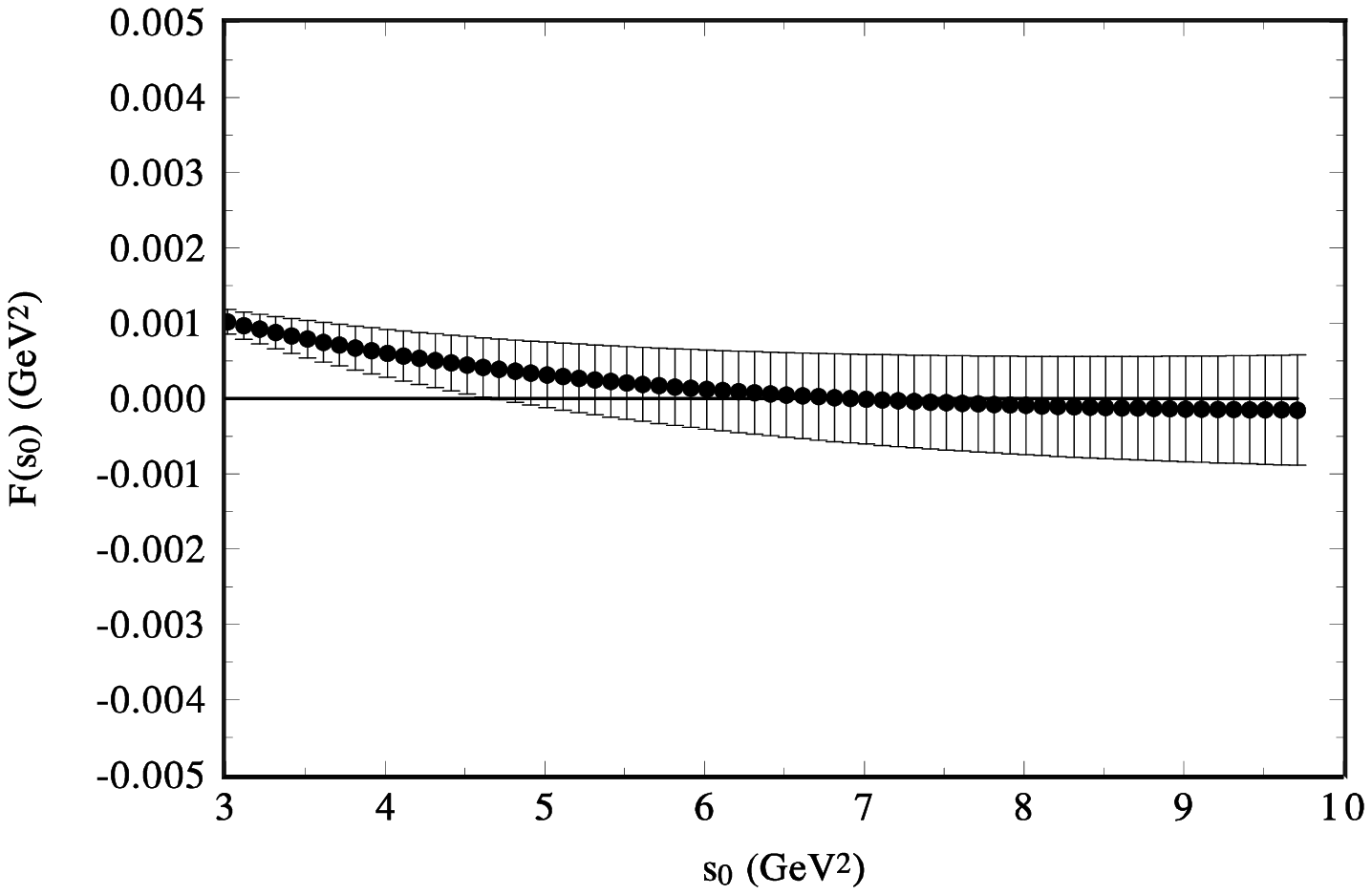}
\end{center}
\caption{$F(s_0)$ is the right hand side of the FESR, Eq.(13), in the vector channel compared with zero, using CIPT with $\Lambda = 397 \, \mbox{MeV}$ ($\alpha_s(M_\tau^2)= 0.353$), $C_2\langle{\cal{O}}_2\rangle =0$, and $C_4\langle{\cal{O}}_4\rangle = 0.05 \, \mbox{GeV}^4$.}
\label{FIG6}
\end{figure}

Next, we turn to the vector channel where the FESR is obtained from Eq.(13) by setting the pion pole contribution to zero. Lacking a driving term that sets the scale, one could analyse the FESR in several ways. We have found that a transparent way is to simply compare the right hand side with zero. The result is shown in Fig.6 for $\Lambda = 397\, \mbox{MeV}$ ($\alpha_s(M_\tau^2)= 0.353$), $C_2\langle{\cal{O}}_2\rangle =0$, and $C_4\langle{\cal{O}}_4\rangle = 0.1 \, \mbox{GeV}^4$; the vertical axis, $F(s_0)$, is the right hand side of the FESR, Eq.(13). The agreement is very good in an unusually broad stability region. Changes in $\Lambda$ and the corresponding dimension $d=4$ condensate are fully compatible with the previous results from the axial-vector channel. Also, the conclusion about $C_2\langle{\cal{O}}_2\rangle$ being  very small remains unchanged. Hence, the compatibility is established between results from the vector and the axial-vector channel, and in turn, their compatibility with PQCD plus no dimension $d=2$ term in the OPE, and a dimension $d=4$ condensate in line with previous independent determinations using different weighted FESR \cite{DS2}-\cite{ASS}.\\

We comment now on the sensitivity of the results to the choice of the cutoff energy $s_1$ which enters in the integration kernel $P(s)$, Eq.(11). The experimental errors increase appreciably as one approaches the kinematical end-point $s = M_\tau^2$. Hence, we have chosen $s_1$ high enough to include as much data as possible, but somewhat below the end-point so that the experimental uncertainties remain under control. In any case, varying $s_1$ in  the reasonable range $s_1 \simeq 2.6 - 3.0 \, \mbox{GeV}^2$ we find that the resulting $f_\pi$ changes (upwards or downwards) only by some $0.4 - 1.3  \,\%$ in the region of stability. The latter remains essentially unchanged in quality.  

Finally, we discuss the error $\Delta(s_0)$ in the FESR, Eq.(13), due to the fact that the data might, in all probability, not be constant above the kinematical end-point. Assuming that PQCD sets in for $s \geq s_1$, we would arrive at the following estimate of this error

\begin{equation}
\Delta(s_0) = \frac{1}{4\pi^2} \left\{ s_0 \,M_2(s_0) - s_1 \,M_2(s_1) - 2\; \frac{\left[s_0^2\, M_4(s_0) - s_1^2 \,M_4(s_1)\right]}{(s_0+s_1)}\right\}.
\end{equation} 

Numerically, we obtain $\Delta(s_0) < 5 \times 10^{-5} \, \mbox{GeV}^2$ inside the stability region, and for $\Lambda = 365 - 380 ,\mbox{MeV}$. This is to be compared with the right hand side of the FESR, Eq. (13), i.e. $ 2\, f_\pi^2 \simeq 1.7\, \times\, 10^{-2}\, \mbox{GeV}^2$. This conclusion is basically insensitive to reasonable changes in $s_1$, due to the fact that $\Delta(s_0)$ is some two orders of magnitude smaller than $2 f_\pi^2$. Alternatively, relaxing the assumption
that PQCD sets in for $s \geq s_1$, we have used $e^{+}  e^{-}$ data \cite{PDG} in the region $s_1 - s_0$, with $s_0$ in the interval  $s_0 = (s_1 - 4) \; \mbox{GeV}^2$, in order to compute the error (deviation from zero) in the vector channel, i.e.
\begin{equation}
\Delta(s_0)|_V = \frac{1}{4 \pi^2} \int_{s_{1}}^{s_{0}} ds \, P(s) \, \frac{1}{\pi}\, Im \Pi(s)|_{e^{+} e^{-} DATA} \;.
\end{equation}
The result is  $\Delta(s_0)|_V \simeq (10^{-5} - 10^{-4}) \mbox{GeV}^2$, i.e. one to two orders of magnitude smaller than  $F(s_0)$ (see Fig. 6). One should notice that for $s_0 > 4\; \mbox{GeV}^2$ the $e^{+}  e^{-}$ data is essentially constant, albeit with very large errors, all the way up till the $J/\psi$ region \cite{PDG} in which case $\Delta(s_0)|_V \equiv 0$ .

\noindent
\section{Conclusions}
\noindent 

In this paper we have introduced a new QCD-FESR involving an integration kernel designed to suppress the hadronic contribution beyond the kinematical end-point of $\tau$-decay, i.e. the region where there is no longer data. While various different kinds of integration kernels in FESR have been used regularly in the past, this particular one is unique in its purpose as well as in the results it produces. In fact, this FESR allows to test the consistency between QCD and the $\tau$-decay vector and  axial-vector hadronic spectral functions. We have found an excellent consistency in a remarkably wide range of energies, to wit. In the axial-vector channel the FESR reproduces, within errors, the  experimental value of the pion decay constant in the broad region $s_0 \simeq 4 - 10 \,\mbox{GeV}^2$, for current values of the QCD strong coupling, and the vacuum condensates. In the vector channel, where the FESR is confronted with zero, agreement is found roughly in  the same stability region.
This consistency would not be at all obvious if non-weighted FESR were to be employed. In fact, as it follows from Fig.1, in this case the saturation of the axial-vector FESR is rather poor. This is one of the central conclusions from this analysis.  And, of course,  the other is that we have shown that by using a suitable integration kernel it is possible to extend a QCD FESR analysis well beyond the kinematical end-point of data. This feature could become very useful in instances where the experimental data is only known in a narrow energy region. 



\begin{thebibliography}{9999}                                                                                             %
\bibitem {KSMT}K. Schilcher and M.D. Tran, Phys. Rev. \textbf{D 29} (1984) 570.

\bibitem{ARGUS} H. Albrecht et al., Z. Phys. \textbf{C 33} (1986) 7.

\bibitem{DSP}C.A. Dominguez and J. Sol\`{a}, Z. Phys. \textbf{C 40} (1988) 63; V. Gimenez, J.A. Pe\~{n}arrocha, and J. Bordes, Phys. Lett. \textbf{B 223} (1989) 245.

\bibitem{QCDSR} For a review see e.g. P. Colangelo and  A. Khodjamirian, in: "At the Frontier of Particle Physics/ Handbook of QCD"', M. Shifman, ed. (World Scientific, Singapore 2001), Vol. 3, 1495-1576.

\bibitem {ALEPH} ALEPH Collaboration, S. Schael \textit{et al.}, Phys. Rept.
\textbf{421} (2005) 191.

\bibitem {OPAL}OPAL Collaboration, K. Ackerstaff \textit{et al.}, Eur. Phys.
J. \textbf{C 7} (1999) 571; G. Abbiendi \textit{et al.}, \textit{ibid.}
\textbf{13} (2000) 197.

\bibitem {Braaten} E. Braaten, S. Narison, and A. Pich,  Nucl. Phys. \textbf{B 373} (1992) 581.

\bibitem{taureview} For a  review see e.g. M. Davier, A. H\"{o}cker, and Z. Zhang, Rev. Mod. Phys. \textbf{78} (2006) 1043; A. Pich, arXiv: 0808.2897. See also O. Cata, M. Golterman, and S. Peris, Phys. Rev. \textbf{D 77} (2008) 093006; {\it ibid.} arXiv:0812.2285.

\bibitem{MALT}  K. Maltman, Phys. Lett. \textbf{B 440} (1998) 367; K. Maltman, and T. Yavin, Phys. Rev. \textbf{D 78} (2008) 094020.

\bibitem{DS1} C.A. Dominguez and K. Schilcher, Physics Letters  \textbf{B 448}  (1999) 93;  Physics Letters \textbf{B 581} (2004) 193.
 
\bibitem {ALPHA1}M. Davier, S. Descotes-Genon, A. H\"{o}cker, B. Malaescu, and Z. Zhang, Eur. Phys. J. \textbf{C 56} (2008) 305.
 
\bibitem {ALPHA2}P. A. Baikov, K. Chetyrkin and J. H. K\"{u}hn, Phys. Rev.
Lett. \textbf{101} (2008), 012002.

\bibitem{DS2} J. Bordes, C.A. Dominguez, J. Pe\~{n}arrocha, and K. Schilcher, J. High Energy Phys. \textbf{0602} (2006) 037.

\bibitem {IZ}G. L. Ioffe and K. N. Zyablyuk, Eur. Phys. J. \textbf{C 27} (2003), 22.

\bibitem {DS3}C.A. Dominguez and K. Schilcher, J. High Energy Phys.
\textbf{0701} (2007), 093.

\bibitem{ASS}  A. Almasy, K. Schilcher, and H. Spiesberger, Eur. Phys. J. \textbf{C 55} (2008) 237.

\bibitem{DS4} C.A. Dominguez and K. Schilcher, Phys. Rev. \textbf{D 61} (2000) 114020.

\bibitem{CAD1} C.A. Dominguez, Phys. Lett. \textbf{B 345} (1995) 291.

\bibitem {NAR} S. Narison, {\it Power corrections to $\alpha_{s}(M_{\tau})$,
$|V_{us}|$ and $\bar{m}_{s}$}, arXiv: 0901.3823, (2009), and references therein.

\bibitem {CIPT}A.A. Pivovarov, Sov. J. Nucl. Phys. \textbf{54} (1991) 676; F. Le Diberder and A. Pich, Phys. Lett. \textbf{ B 286} (1992) 147.

\bibitem {CHET1}K.G. Chetyrkin, B.A. Kniehl, M.Steinhauser, Phys. Rev. Lett. \textbf{79} (1997) 2184.

\bibitem{PDG} Particle Data Group. C. Amsler {\it et al.}, Phys. Lett. B 667, 1, (2008).
\end{thebibliography}
\end{document}